\documentstyle[a4,12pt]{article}
\begin{document}
\begin{flushright}
LPT-ORSAY 99/25 \\
hep-th/9905012
\end{flushright}
\vskip 1cm
\begin{center}
{\Large \bf Non-conventional cosmology from a brane-universe}\\
\vskip 2cm
{Pierre Bin\'etruy$^1$, C\'edric Deffayet$^1$, David Langlois$^2$\\ 
\vskip 1cm 
$^1$ LPT\footnote{Unit\'e mixte de recherche UMR n$^o$ 8627.}, 
Universit\'e Paris-XI, B\^atiment 210, 
F-91405 Orsay Cedex, France;\\
$^2$ D\'epartement d'Astrophysique Relativiste et de Cosmologie,
C.N.R.S.,\\ Observatoire de Paris, 92195, Meudon, France.} \\

\vskip 1.6cm 
{\bf Abstract.}  
\vskip 1cm
\end{center} 

{ We consider ``brane-universes'', where matter is confined 
to  four-dimensional hypersurfaces (three-branes) whereas  one extra
compact dimension is felt by gravity only. We show that the cosmology of 
such branes is definitely different from standard cosmology and
identify the reasons behind this difference. We give a new class of 
exact solutions with a constant five-dimensional radius and 
cosmologically evolving brane.
We discuss various consequences. }

\def\beq{\begin{equation}}
\def\eeq{\end{equation}}
\def\d{\delta}
\def\fourG{{{}^{(4)}G}}
\def\4R{{{}^{(4)}R}}
\def\H{{\cal H}}
\def\K5{{\kappa_{(5)}}}

\newcommand{\A}{A}
\newcommand{\B}{B}
\newcommand{\mmu}{\mu}
\newcommand{\mnu}{\nu}
\newcommand{\ii}{i}
\newcommand{\jj}{j}
\newcommand{\jl}{[}
\newcommand{\jr}{]}
\newcommand{\ml}{\sharp}
\newcommand{\mr}{\sharp}

\newcommand{\da}{\dot{a}}
\newcommand{\db}{\dot{b}}
\newcommand{\dn}{\dot{n}}
\newcommand{\dda}{\ddot{a}}
\newcommand{\ddb}{\ddot{b}}
\newcommand{\ddn}{\ddot{n}}
\newcommand{\pa}{a^{\prime}}
\newcommand{\pb}{b^{\prime}}
\newcommand{\pn}{n^{\prime}}
\newcommand{\ppa}{a^{\prime \prime}}
\newcommand{\ppb}{b^{\prime \prime}}
\newcommand{\ppn}{n^{\prime \prime}}
\newcommand{\fda}{\frac{\da}{a}}
\newcommand{\fdb}{\frac{\db}{b}}
\newcommand{\fdn}{\frac{\dn}{n}}
\newcommand{\fdda}{\frac{\dda}{a}}
\newcommand{\fddb}{\frac{\ddb}{b}}
\newcommand{\fddn}{\frac{\ddn}{n}}
\newcommand{\fpa}{\frac{\pa}{a}}
\newcommand{\fpb}{\frac{\pb}{b}}
\newcommand{\fpn}{\frac{\pn}{n}}
\newcommand{\fppa}{\frac{\ppa}{a}}
\newcommand{\fppb}{\frac{\ppb}{b}}
\newcommand{\fppn}{\frac{\ppn}{n}}

\newcommand{\dA}{\dot{A_0}}
\newcommand{\dB}{\dot{B_0}}
\newcommand{\fdA}{\frac{\dA}{A_0}}
\newcommand{\fdB}{\frac{\dB}{B_0}}

\section{Introduction}
There has been recently renewed activity in the domain of cosmology with 
extra dimensions. Historically, investigations on the possibility of 
extra dimensions began with Kaluza-Klein type theories (see e.g. \cite{bl87} 
for a review) and were then  revived with the advent of high energy physics 
theories, such as string and superstring theories, where the existence of 
additional dimensions seemed necessary to ensure a non-anomalous quantum 
behaviour. However, these extra dimensions, since they remained undetected in 
experiments, were constrained to be very small. 

Recently, it has been suggested \cite{add98} that additional dimensions
could have a very 
distinct nature from our familiar dimensions, in the sense that ordinary 
matter would be confined to our four-dimensional ``universe'' while gravity 
would ``live'' in the whole extended spacetime. In other words, our 
four-dimensional universe would be a three-brane living in a 
$(4+n)$-dimensional spacetime ($n$ being the number of additional dimensions).
One of the advantages of this  picture would be to relax the strong 
constraints on the size of extra dimensions and to allow for large additional 
dimensions, with the associated property of a low fundamental Planck scale, 
leading to many phenomenological consequences. 

A strong motivation for considering such models comes from  strongly
coupled string theories \cite{hw96},\cite{aahd98}. For example, the
strongly coupled $E_8 \times E_8$ heterotic string theory  is believed to be an
11-dimensional theory, the field theory limit of which was described
by Ho\v{r}ava and  Witten \cite{hw96}. The spacetime manifold includes a
compact dimension with an orbifold structure. Matter is confined on the two
10-dimensional hypersurfaces (9-branes) which can be seen as forming the 
boundaries of this spacetime.

For all such theories, an essential issue concerns the cosmological evolution
of our universe. Depending on the mass scales associated, this might indeed
be the only way to test such models. Several recent works have examined this
question \cite{kko97} \cite{low98}, with a special emphasis on inflation
\cite{l99}-\cite{ko99}. The purpose of the present work is
to reconsider the cosmological evolution of a brane-like universe. 
Our main  result is that, contrarily to what has been implicitly
assumed in  previous works, brane cosmology leads to
Friedmann-like equations very  different from the standard ones.

Since this area of research lies at the interface between high energy physics 
and general relativity, we have tried, as much as possible, to present our 
results from both viewpoints. We have restricted our analysis to a single 
additional dimension, leaving the more general case of several 
extra dimensions for future work, but we expect that our main result, {\it i.e.} the 
non conventional nature of the cosmological dynamics for a brane-universe, 
will still hold qualitatively. The plan of this paper is the following. 
In section 2, we present the general framework for our five-dimensional 
effective theory. In section 3, we compute 
the corresponding five-dimensional Einstein equations and study the
conditions imposed locally by the presence of a brane. In section 4,
we adopt a geometrical point of view and rewrite 
the Einstein equations in the Gauss-Codacci form. In section 5, we
adopt a global point of view and show that this imposes some
relations between the energy-momentum of matter in our brane-universe
and that of matter in the rest of the five-dimensional spacetime. We identify
a class of exact solutions which show explicitly the non-conventional
dynamics discussed in this paper; this allows us to discuss the
behaviour of the zero modes in  a Kaluza-Klein type 
approach. Finally, section 6 discusses  the  consequences of our 
unconventional cosmological equations.

\section{Brane theory in five-dimensional spacetime}

The cosmological consequences of having $n$ extra compact dimensions unseen by
ordinary matter depend on the  number $n$ of such dimensions
\cite{kl98,low99}. In this article, we will restrict our attention to the
case of a single such dimension.  One motivation is that this corresponds
to the case  studied by  Ho\v{r}ava and  Witten \cite{hw96} and this
explicit example may provide some insight and some explicit tests of the 
properties discussed on more general terms. We will actually
consider for illustrative purposes  somewhat simpler theories, obtained by
compactification to five dimensions. ``Our''
four-dimensional Universe will be assimilated to a four-dimensional 
hypersurface (three-brane).

The theory we are thus considering is five-dimensional and has an action 
of the form  
\beq 
S_{(5)}=-\frac{1}{2\kappa^2_{(5)}} \int d^5x
\sqrt{-\tilde{g}} \tilde{R} +\int d^5x \sqrt{-\tilde{g}}{\cal L}_m,
\label{action} 
\eeq 
where the first term corresponds to the Einstein-Hilbert
action in five  dimensions for a five-dimensional metric ${\tilde g}_{AB}$,
$\tilde{R}$  being the five-dimensional scalar curvature  for this metric,
and the second term corresponds to the matter content. Our notational
conventions will be as follows: upper case Latin letters denote
five-dimensional indices  $(0,1,2,3,5)$; lower case Greek indices run over
the four conventional  dimensions $(0,1,2,3)$ whereas lower case Latin
indices run over ordinary  spatial dimensions $(1,2,3)$. To avoid confusion,
some five-dimensional fields  (like $\tilde{g}_{AB}$) will be tilted in
contrast with four-dimensional  fields. The signature of $\tilde{g}_{AB}$ is
$(-++++)$. $\nabla_A$ will denote the covariant derivative associated 
with the metric $\tilde{g}_{AB}$.
 The constant $\kappa_{(5)}$ in the action (\ref{action}) is directly related to
the  five-dimensional Newton's constant $G_{(5)}$ and the 
 five-dimensional reduced Planck mass ${M}_{(5)}$,
by the relations
\beq
 \kappa^2_{(5)} = 8 \pi G_{(5)} = {M}_{(5)}^{-3}.
\eeq

We shall consider  five-dimensional spacetime metrics of the form 
\begin{equation}
ds^{2}=\tilde{g}_{AB}dx^A dx^B = 
g_{\mmu \mnu} dx^{\mmu}dx^{\mnu} + b^{2}dy^{2} 
\end{equation}
where $y$ is the coordinate of the fifth dimension which we assume to 
be compact; its range interval  $I$ is
chosen to be  $-1/2 \le y \le 1/2$, with the two endpoints
of the interval identified. 

Inspired by the compactification of  Ho\v{r}ava-Witten theory, we will
sometimes  impose a symmetry $y \rightarrow -y$. 
In the specific model of Ho\v{r}ava-Witten \cite{hw96},  
points obtained from one
another by this symmetry are identified. 
The relevant interval is then $0 \le y \le
1/2$ and there are two three-branes which form the boundary of spacetime,
respectively at $y=0$ and $y=1/2$.

Throughout this article, we will focus our attention on 
the hypersurface defined by $y=0$, which we  identify
with the  world volume of the brane that  forms our universe.
 Since we are interested in cosmological solutions, we take a metric of the form
\begin{equation}
ds^{2}=-n^{2}(\tau,y) d\tau^{2}+a^{2}(\tau,y)\delta_{ij}dx^{i}dx^{j}+b^{2}(\tau,y)dy^{2},
\label{metric}
\end{equation}
where we have 
assumed for simplicity a flat metric in the ordinary spatial dimensions.  
Let us turn to the matter part of the theory. 
We shall distinguish two kinds of  sources: fields confined to the world-volume
of our brane-universe  and fields living in the bulk spacetime.   
The energy-momentum tensor, derived from the matter part of the 
action (\ref{action}), will thus be decomposed into 
\begin{equation}
\tilde{T}^\A_{\quad \B}  =  \check{T}^\A_{\quad \B}|_{_{\rm bulk}} 
+ T^\A_{\quad \B}|_{_{\rm brane}},
\end{equation}
where $\tilde{T}^\A_{\quad \B}|_{_{\rm bulk}}$ is the energy momentum tensor 
of the bulk matter (and possibly other branes), which we do not need to 
specify here, and the second 
term $T^\A_{\quad \B}|_{_{\rm brane}}$ corresponds to the matter content
in the 
brane $(y=0)$. The latter can be expressed quite generally in the form
\begin{equation}
T^\A_{\quad \B}|_{_{\rm brane}}= \frac{\delta (y)}{b} \mbox{diag} 
\left(-\rho,p,p,p,0 \right), 
\label{source}
\end{equation}
where the energy density $\rho$ and pressure $p$ are independent of the 
position in the brane in order to recover a homogeneous cosmology in the 
brane.
Here we have assumed an idealized situation where the physical brane is 
assumed to be infinitely thin. 
Rigorously,  the physical brane will have a width in the fifth dimension 
which is usually assumed to be given by the fundamental scale of the 
underlying theory.
In our case, the thin-brane approximation will thus be valid when the 
energy scales at which we consider the theory
are much smaller than the fundamental scale (which is anyway required if one
describes gravity classically).

Let us emphasize at this stage that the standard way to probe the metric of
the  universe on cosmological scales is through the kinematics of the 
PPSM (Particle Physics Standard Model) fields.
If one assumes that the PPSM fields live on the brane, they 
will be sensitive only to the {\it induced} metric 
(given here by ${g}_{\mmu \mnu}(\tau,0))$. 
 For example, in the case of  redshift  measurements, 
the redshift of a given distant galaxy is 
$\lambda_0 / \lambda_1 -1
= (a(\tau_0,0) / a(\tau_1,0))-1$ 
(where $\tau_1$ and $\tau_0$ are respectively the time of emission
and the time of reception). 
Similarly,   the equation of conservation for the  energy-momentum in the
brane, which will be given below in (\ref{energy}), 
depends only on the metric in $y=0$. The four-dimensional effective cosmology
for our world 
is thus given by the equations relating the induced metric 
$g_{\mmu \mnu}(\tau,0)$ to the sources. The determination of the 
dynamics of this induced metric will be the object of the 
 next two sections, whose essential feature will be to solve locally 
the Einstein equations in the vicinity of the brane.

\section{Effective brane cosmology from Einstein equations}

The dynamics of the five-dimensional geometry, thus including the 
brane geometry, is governed  by the 
five-dimensional  Einstein equations, which can be derived from the action 
(\ref{action})
and take the usual form
\beq
{\tilde G}_{\A\B}=\kappa^2_{(5)} \tilde{T}_{\A\B}. \label{einstein}
\eeq
Inserting the ansatz (\ref{metric}) for the metric, the non-vanishing 
components of the Einstein tensor 
${\tilde G}_{\A\B}$ are found to be 
\begin{eqnarray}
{\tilde G}_{00} &=& 3\left\{ \fda \left( \fda+ \fdb \right) - \frac{n^2}{b^2} 
\left(\fppa + \fpa \left( \fpa - \fpb \right) \right)  \right\}, 
\label{ein00} \\
 {\tilde G}_{\ii\jj} &=& 
\frac{a^2}{b^2} \delta_{ij}\left\{\fpa
\left(\fpa+2\fpn\right)-\fpb\left(\fpn+2\fpa\right)
+2\fppa+\fppn\right\} 
\nonumber \\
& &+\frac{a^2}{n^2} \delta_{ij} \left\{ \fda \left(-\fda+2\fdn\right)-2\fdda
+ \fdb \left(-2\fda + \fdn \right) - \fddb \right\},
\label{einij} \\
{\tilde G}_{05} &=&  3\left(\fpn \fda + \fpa \fdb - \frac{\dot{a}^{\prime}}{a}
 \right),
\label{ein05} \\
{\tilde G}_{55} &=& 3\left\{ \fpa \left(\fpa+\fpn \right) - \frac{b^2}{n^2} 
\left(\fda \left(\fda-\fdn \right) + \fdda\right) \right\}.
\label{ein55} 
\end{eqnarray} 
In the above expressions, a prime stands for a derivative with respect to
 $y$, and a 
dot for a derivative with respect to $\tau$. 

From the Bianchi identity $\nabla_{\A}{\tilde G}^{A}_{\quad \B}=0$ 
we can obtain, using 
the Einstein equations (\ref{einstein}), an equation of conservation of the form
\begin{equation}
 \dot{\rho} + 3(p+\rho) \frac{\dot{a_0}}{a_0}  =0,
\label{energy}
\end{equation}
which is the usual four-dimensional equation of conservation for the 
energy density in cosmology ($a_0$ is the value of $a$ on the brane). 
It will be shown in the next section how 
one recovers more generally the conservation of the stress-energy-momentum 
tensor of the brane.

We now seek  a solution of Einstein's  equations (\ref{einstein}) 
in the vicinity of $y=0$. 
In order to have a well defined geometry, the metric is required 
 to be  continuous across the brane localized in $y=0$. However its 
 derivatives with respect to $y$  can   be discontinuous 
in $y=0$. This will entail the existence of a Dirac delta function
in the second derivatives of the metric with respect to $y$. The resulting 
terms with a  delta function appearing in  the Einstein tensor
(in its components (\ref{ein00}) and (\ref{einij})) must  be matched 
with the distributional components from the stress-energy tensor 
in order to satisfy Einstein's equations. 
For $a$, one has for example
\begin{equation}
a^{\prime \prime}  =  \widehat{a^{\prime \prime}} 
+ \jl a^\prime\jr \delta (y), \label{asec}
\end{equation}
where $\widehat{ a^{\prime \prime} } $ is the non-distributional 
part of the double derivative of $a$ (the standard
derivative) , and $[a^\prime]$ is the {\it jump} in the first derivative across $y=0$, defined by
\begin{equation}
[a^\prime] =a^\prime(0^+)-a^\prime(0^-). \label{jump}
\end{equation}
For future use we also define the {\it mean value} of a given function $f$ 
across $y=0$ by
\begin{equation}
\ml f \mr =\frac{f(0^+)+f(0^-)}{2}. 
\end{equation}
 Matching the Dirac delta functions in the components 
 (\ref{ein00}) and (\ref{einij}) of the Einstein tensor with those 
of the brane stress-energy tensor (\ref{source}), one obtains the 
following relations,
\begin{eqnarray}
\frac{\jl a^\prime \jr}{a_0 b_0}&=&-\frac{\K5^2}{3}\rho, \label{aarho}\\
\frac{\jl n^\prime \jr}{n_0 b_0}&=&\frac{\K5^2}{3}  \left( 3p + 2\rho \right), 
\label{nnrho}
\end{eqnarray}
where the subscript $0$ for $a,b,n$ means that these functions are 
taken in $y=0$.

If we take the {\it jump} of the component $(0,5)$ of Einstein equations, 
and use the above relations (\ref{aarho}) and 
(\ref{nnrho}),  we recover the  energy conservation equation 
(\ref{energy}).
Let us now take the {\it jump} of the component $(5,5)$ of 
Einstein's equations. Using  (\ref{aarho}-\ref{nnrho}), we obtain
\begin{equation}
\frac{\ml a^\prime \mr}{a_0} p =\frac{1}{3} \rho 
\frac {\ml n^\prime \mr}{n_0}, \label{mean}
\end{equation}
whereas the {\it mean value} of the same equation gives
\begin{equation}
{1\over n_0^2}\left(\frac{\dot{a_0}^2}{a_0^2}
-\frac{\dot{a_0}}{a_0} \frac{\dot{n_0}}{n_0}
+\frac{\ddot{a_0}}{a_0}\right) = 
-\frac{\K5^4}{36}\rho(\rho+3p)  
+\frac{\left(\ml a^\prime \mr \right)^2}{a_0^2 b_0^2}
\left( 1+\frac{3p}{\rho} \right) -  {\K5^2 \check T_{55}\over 3 b_0^2},
\label{FriedFried}
\end{equation}
where we recall that $\check T_{AB}$ is the bulk energy-momentum tensor.

If we restrict our analysis to solutions that are left invariant by the
  symmetry $y \rightarrow -y$ (as in Horava-Witten supergravity), then things
simplify considerably since  $a$ is now  a function of 
$|y|$, and $\ml a^{\prime}\mr = 0$. 
The above equation then simplifies to the form
\begin{equation}
  \frac{\dot{a_0}^2}{a_0^2}+\frac{\ddot{a_0}}{a_0}=
-\frac{\K5^4}{36}\rho(\rho+3p)-{\K5^2 \check T_{55}\over 3 b_0^2},
\label{Fried}
\end{equation}
where we have chosen the time $\tau$, without any loss of generality, so
that $n_0=1$ (this corresponds to the usual cosmic time in standard cosmology).
This equation is one of  the main results of the present work and must 
be interpreted as a new Friedmann-type equation. 
It  must be emphasized that this equation (as well as 
(\ref{FriedFried})) is fundamentally different from the 
standard cosmological Friedmann equation in the sense that the square 
of the Hubble parameter $H$   {\it depends quadratically} on the 
cosmological ({\it i.e.} the brane) energy density, in contrast with 
the usual linear dependence.\footnote{We will show later that the
same quadratic dependence would hold, to leading order, if we were to
define the Hubble parameter through the zero modes of the metric, in a
Kaluza-Klein-like way.}

Let us note from (\ref{Fried}) that 
the universe dynamics will be dominated by  the brane
energy density rather than the bulk pressure (assumed to be of the order of 
the bulk energy density)  when the condition 
 \beq
\rho_{bulk} \ll {\rho_{brane}^2 \over M^3_{(5)}} \label{dombrane}
\eeq
is satisfied. 

Assuming that (\ref{dombrane}) is verified, one important  consequence 
is that, for the {\it same 
matter content} in our four-dimensional universe, {\it the scale factor 
evolution will be different} from the standard one. 
Indeed, let us consider an equation of state 
of the form $p=w\rho$, with $w$ constant, which includes the  main 
effective equations of state used in cosmology, namely a radiation gas 
(with $w=1/3$), a non-relativistic gas  (with 
$w\simeq 0$) and also a cosmological constant (with $w=-1$). 
The energy conservation law (\ref{energy}), unchanged with 
respect to the standard case, yields the usual relation between the 
energy density and the scale factor, 
\beq
\rho \propto a_0^{-3(1+w)}.
\eeq
However, when one now looks for power-law solutions for the evolution of
the scale factor, {\it i.e.} such that 
\beq
a_0(t)\propto t^q,
\eeq
then the new Friedmann-type equation (\ref{Fried}), neglecting the 
bulk term, leads to 
\beq
q={1\over 3(1+w)},
\eeq
which must be contrasted with the standard cosmological solutions for which 
$q_{standard}=2/(3(1+w))$ (the particular case $w=-1$ corresponding 
to a cosmological constant will be addressed below).
The new cosmological equation thus leads typically 
to slower evolution. 

Another essential feature of the equation (\ref{FriedFried}) is that the 
effective four-dimensional Newton's constant does not appear in it. The 
dynamics of the universe, from this equation, depends only on the 
fundamental five-dimensional Newton's constant.

This result seems in contradiction with most of the recent litterature 
on brane cosmology (see in particular \cite{low99} \cite{adkm99}), where 
the standard Friedmann equations are recovered as an effective four-dimensional
description. There exist, however, some hints of unconventional cosmology
(see \cite{kl98} and the ``non-linear'' regime of \cite{low99}) but they 
are presented as valid only for high energies, whereas our results
indicate that they are the rule as far as the five-dimensional description 
makes sense.

 In what follows, we will try to get  a better  understanding of the 
above non standard Friedmann equations, from different viewpoints
including the derivation of a new class of exact solutions.

\section{Gauss-Codacci equations}
The purpose of this section is to give a geometrical  analysis of a 
three-brane living in a five-dimensional spacetime. It will be useful to resort to
a well-known technique in general relativity, which consists in decomposing the Einstein equations 
into a part tangential to the surface under consideration, a part normal to 
it and finally a mixed part (since Einstein equations are tensorial 
of order two). This leads in particular to the so-called Gauss-Codacci 
equations. This technique was used to study thin shells in general relativity 
\cite{israel66} and is particular fit to study the very analogous problem 
of branes.

Let us first introduce the unit vector field $n^A$ 
normal to the three-brane worldsheet. Note that this vector field is spacelike,
{\it i.e.} 
\beq
g_{AB}n^A n^B=1. \label{gc1}
\eeq
 The induced metric on the brane worldsheet will thus 
be defined by
\beq
h_{AB}=g_{AB}-n_An_B. \label{gc2}
\eeq
A very useful quantity  is the extrinsic curvature tensor $K_{AB}$ (also 
called second fundamental form), which is defined by the expression 
\beq
K_{AB}=h_A^C\nabla_C n_B, \label{gc3}
\eeq
and can be interpreted as representing the ``bending'' of the brane 
worlsheet in the five-dimensional spacetime. As will be shown below, the 
non-conventional cosmological equations which we shall find are a direct 
consequence of the fact 
that $K_{AB}$ is  non-vanishing in the presence 
of matter confined to the brane, which means that the five-dimensional 
metric depends  necessarily on the fifth dimension, in contrast with 
the usual assumption in the Kaluza-Klein approach. 

The five-dimensional Einstein tensor ${\tilde G}_{AB}$ can then be decomposed 
into the 
following projections,
\beq
{\tilde G}_{AB}n^A n^B =-{1\over 2}\4R+{1\over 2}\left(K^2-K_{AB}K^{AB}\right),
\label{gc4}
\eeq
\beq
n^A {\tilde G}_{AB}h^B_C=D_A K^A_C-D_C K, \label{gc5}
\eeq
\begin{eqnarray}
{\tilde G}_{AB} h^A_C h^B_D&=&{\fourG}_{CD} -K K_{CD}-n^E\nabla_E K_{CD} +D_C a_D
-2n_{(C}K_{D)E}a^E- a_C a_D \cr
&+& \left({1\over 2}K^2 +{1\over 2}K_{AB}K^{AB}+n^A\nabla_A K-\nabla_B a^B
\right)h_{CD}, \label{gc6}
\end{eqnarray}
where $a^B=n^C\nabla_C n^B$ is the ``acceleration" vector field, 
$D_A$ is the covariant derivative associated with the induced metric
$h_{AB}$, $K=g^{AB}K_{AB}$ is the trace of the extrinsic curvature
tensor, and the parenthesis around two indices denote symmetrisation 
 with weight $1/2$.
The  above equations  directly follow from  the  Gauss-Codacci relations 
(see e.g. \cite{wald}). Of course, these equations are simply a rewriting 
of the equivalent relations (\ref{ein00})-(\ref{ein55}), although more general in the sense that we 
have not specified any ansatz for the metric. The components 
(\ref{ein00})-(\ref{ein55}) can be 
recovered directly from the above relations, by noting that, in the case 
of a metric of the form (\ref{metric}), one finds $n^A=\{0,0,0,0,1/b\}$, 
$a_B=\{-\dot b/b,0,0,0,0\}$ and 
$K^A_B=\mbox{diag}\{ n'/ (nb),\d^i_j a'/ (ab), 0\}$.

Let us now examine the above equations from a distributional point of view.
The stress-energy tensor (\ref{source})
associated with the brane can be rewritten in the form 
\beq
T^\A_{\quad \B}|_{_{\rm brane}}=S^A_{\, B}{\d(y)\over b}. \label{gc7}
\eeq
The bulk stress-energy tensor is regular in the vicinity of $y=0$ and is 
$y$-dependent. The 
metric is assumed, as usual, to be continuous. Thus,
only second derivatives of the metric 
with respect to the normal coordinate $y$ can contain a Dirac distribution,
whereas 
the extrinsic curvature can at most contain a Heavyside distribution.
It is clear from the Einstein tensor decomposition (\ref{gc4}-\ref{gc6})
 given above 
that only the term $n^E\nabla_E K^\mu_\nu -\delta^\mu_\nu n^A\nabla_A K
=\partial_y\left(K^\mu_{\, \nu}
-K\delta ^\mu_{\, \nu}\right)/b$ contains second derivatives with respect to 
$y$. This implies the following equation
\beq
\left[K^\mu_{\, \nu}
-K\delta ^\mu_{\, \nu}\right]=-\kappa_{(5)}^2 S^\mu_{\, \nu},\label{gc8}
\eeq
where the brackets here denote the discontinuity between the two sides of the 
brane, as defined in (\ref{jump}). The above relation is known to general 
relativists as Israel's junction condition \cite{israel66}.
It can be  rewritten  in the form 
\beq
\left[ K_{\mu\nu} \right]=
-\kappa_{(5)}^2 \left(S_{\mu\nu}-{1\over 3}Sg_{\mu\nu}\right),
\label{gc10}
\eeq
where $S\equiv S_{\mu\nu} g^{\mu\nu}$ is the trace of $S_{\mu\nu}$. 
This equation corresponds to the equations (\ref{aarho})-(\ref{nnrho}) 
of the previous section.

If one assumes  in addition a symmetry  
$y\rightarrow -y$ for the metric 
then the right and left limits of the extrinsic curvature tensor 
are necessarily opposite,
{\it i.e.}
\beq
\left[ K_{\mu\nu} \right] =2  K_{\mu\nu}^+\equiv 2\bar  K_{\mu\nu}.  
\label{gc9}
\eeq
 Then (\ref{gc10}) means that the extrinsic curvature of the brane is 
{\it completely determined by its matter content}. 
Note also that, except for the term with second derivatives, 
all the other terms in $\tilde G^5_{\, 5}$ and $\tilde G^\mu_{\, \nu}$ 
involving 
the extrinsic
curvature are {\it quadratic} in $K_{\mu\nu}$: this is compatible with 
having the same bulk matter on the two sides of the brane  if one 
has the symmetry $K_{\mu\nu}^+=-K_{\mu\nu}^-$, which is the case here.

Inserting the explicit expression of $\bar K_{\mu\nu}$ into (\ref{gc4}) 
leads to the 
following expression for the scalar four-dimensional curvature:
\beq
\4R={\kappa_{(5)}^4\over 4}\left({1\over 3}S^2-S_{\mu\nu}S^{\mu\nu}\right)
-2\kappa_{(5)}^2\check T^5_5. \label{newfriedmann}
\eeq
We thus recover the analogous of equation (\ref{Fried}).

Let us now consider the mixed part of Einstein's equations, 
{\it i.e.} $\tilde G^5_{\, \mu}$, 
corresponding to Eq. (\ref{gc5}). 
Around the brane, this equation implies
\beq
D_\lambda \bar K^\lambda_\mu-D_\mu\bar K=0,
\eeq
which, rewriting $\bar K_{\mu\nu}$ in terms of the stress-energy tensor 
$S_{\mu\nu}$, according to (\ref{gc10}), yields simply the usual four-dimensional 
stress-energy conservation law, {\it i.e.} 
\beq
D_\lambda S^\lambda_\mu=0.
\eeq

To conclude this section, let us stress that the derivation of equation 
(\ref{newfriedmann}) is more general than in the previous section because
we did not need to assume homogeneity and isotropy in the brane. Note 
also that this approach is local and nothing was said or needed 
about the global 
nature of the five-dimensional spacetime.

\section{Global solutions}
In the previous sections, we have focused our attention to the 
behaviour of the geometry  on our brane-universe and in its neighbourhood. 
It is extremely interesting, although difficult, 
 to obtain some exact solutions of the 
global five-dimensional spacetime. We have managed to find a class 
of global solutions with a constant five-dimensional radius and valid for 
a general equation of state for the matter in our brane-universe. Before
presenting them we discuss the general properties of such solutions. 
We  then discuss the  implications  in a Kaluza-Klein  perspective. 

\subsection{General properties of the solutions}

An important constraint comes from the fact that 
the  solutions to Einstein's equations must be seen  as defined {\it globally} 
throughout spacetime, {\it i.e.} for $-1/2 \leq y \leq 1/2$. This implies in 
particular that, from the global point of view that we now adopt, the
second  derivative of $a$ necessarily satisfies an equation more
complicated than (\ref{asec}).   

Indeed there is no Green function $G(y)$ on a circle,
solution of $G^{\prime \prime} = \delta(y)$ \footnote{As is well known 
from electromagnetism, one cannot put a non zero charge alone in a
compact manifold with no boundary.}. On the other hand, one can solve 
$G ^{\prime \prime} = \delta(y) - \delta(y-y_{0})$  or $G^{\prime
 \prime} = \delta(y) - 1$. This means that,  in order to make a metric 
solution of the Einstein equations
 {\it globally defined}, one must introduce another brane (or matter in
the bulk\footnote{This would correspond to the presence of sources
distributed over the entire bulk with an energy-momentum tensor of the
same order as that of the brane. The fundamental three-form of
eleven-dimensional supergravity would provide interesting candidates for
such sources: one could mention the Ramond-Ramond scalar. Studying this
possibility goes beyond the scope of this paper.}). For
simplicity, we will put this other brane at $y=1/2$.
We thus consider an additional term in the stress-energy tensor 
of the form 
\begin{equation}
T^{A}_{*\quad \B}|_{_{\rm brane_*}}= b^{-1}\delta \left( y-1/2 \right) 
 \mbox{diag} \left(-\rho_*,p_*,p_*,p_*,0 \right),
\end{equation}
where a star will be used to denote quantities associated with the second
 brane. 

Keeping only the distributional part for the metric, the second
derivative  of $a$ then satisfies the following differential 
equation:
\begin{equation}
a'' = \jl a' \jr_0 \, \left( \delta(y) - \delta(y-1/2) \right) +
\left(  \jl a' \jr_0 + \jl a'\jr_{1/2}\right) \, \left( \delta(y-1/2) -1
\right)  , \label{aseconde}
\end{equation}
where $\jl a' \jr_{0}$, resp.  $\jl a' \jr_{1/2}$, is the jump of $a'$ on
the  first, resp. second, brane given as in  (\ref{aarho}) by
\begin{eqnarray}
\frac{\jl a^\prime \jr_{0}}{a_{0} b_{0}} &=& -\frac{\K5^2}{3}\rho,
\label{aarho0} \\
\frac{\jl a^\prime \jr_{1/2}}{a_{1/2} b_{1/2}} &=& -\frac{\K5^2}{3}\rho_*.
\label{aarho1/2}
\end{eqnarray}
Integrating (\ref{aseconde})  over $y$ yields the following solution for $a$:
\begin{equation}
a= a_0 + \left( {1 \over 2} |y| - {1 \over 2} y^2 \right) \jl a' \jr_0
- {1\over 2} y^2 \jl a' \jr_{1/2}, \label{aglobal}
\end{equation}
where we recognize the quadratic behaviour found by Lukas , Ovrut and
Waldram \cite{low99}. A similar expression is obtained for $n$. Allowing
a linear dependence in $y$ for the function $b$, we write: 
\begin{equation}
b = b_0 +2 |y| (b_{1/2} - b_0), \label{blinear}
\end{equation}
where $b_0$ is assumed to be constant in time.
Then, plugging these functions in the (0,0) Einstein equation on the
brane at $y=0$, one obtains from (\ref{ein00}):
\begin{equation}
{\dot a_0^2 \over a_0^2} = {n_0^2 \over b_0^2} \left\{ -{\jl a' \jr_{1/2}
\over a_0 } -{b_{1/2} \over b_0} {\jl a' \jr_{0} \over a_0 }
+{1 \over 4} { \jl a' \jr_{0}^2 \over a_0^2 } \right\}. \label{01/2}
\end{equation}
We have seen in section 3 that the (5,5) component of Einstein's
equations imposes $\dot a_0 / a_0$ to be of the order  
of the energy density $\rho$. Thus the leading terms in 
(\ref{01/2}) must cancel, which leads to the equation:
\begin{equation}
{ \jl a' \jr_0 \over b_0} = -{ \jl a' \jr_{1/2} \over b_{1/2}},
\label{aprimconstr}
\end{equation}
and similarly for $n$, using (\ref{einij}).
In other words, using (\ref{aarho0}) and (\ref{aarho1/2}), we must impose:
\begin{eqnarray}
\rho a_0 &=& - \rho_* a_{1/2}, \label{rhoconstr} \\
(2 \rho + 3 p) n_0 &=& - (2 \rho_* + 3 p_*) n_{1/2}. \label{pconstr} 
\end{eqnarray}
Hence the matter on one brane is constrained by the matter on the
other.
 This is related to the fact that most of the information (even
some global one) on the
solutions of the Einstein equations may be obtained locally, as shown in
section 3. The  constraints between the two branes obtained above can probably be
seen as a particular example of ``topological constraints", which impose
specific restrictions on the  distribution of localized matter in a space that
contains compact dimensions and which can be found  in many different
contexts (D-branes, orientifolds or topological defects).

It turns out that, if we assume $b_{1/2}$ to be constant in time, in 
addition to $b_0$, the metric given by (\ref{aglobal}), the similar expression for
$n$ and (\ref{blinear}), yields an exact solution of Einstein's equations
(the time evolution for $a_0$ being given by (\ref{01/2}), and that of 
$\rho$ by the energy conservation equation on the brane). 
However, it can be shown that this class of solutions is simply 
a rewriting (up to a change of the coordinate $y$) of the class
of exact solutions, satisfying the condition
$b_0=b_{1/2}$, exhibited explicitly below.


\subsection{Exact cosmological brane solution}
Let us look for the simplest solutions one can envisage, namely 
solutions linear in $|y|$ for $a$ and $n$, of the form 
\begin{eqnarray}
a&=&a_0(t) \; (1+\lambda |y|), \label{exactsolution}\\
n&=&n_0(t) \; (1+\mu |y|), \\
b&=& b_0, 
\label{linearmetric}
\end{eqnarray}
where $b_0$ is assumed to be constant in time. $\lambda$ 
and $\mu$ are in general functions of time and depend directly on the 
matter content of the brane according to the equations (\ref{aarho}) and 
(\ref{nnrho}), 
\begin{equation}
\lambda = -\frac{\kappa^2_{(5)}}{6} b_0\rho, \qquad
\mu=\frac{\kappa^2_{(5)}}{2}\left(w+{2\over 3}\right)b_0\rho. 
\label{lambdarho}
\end{equation}
We have introduced $w \equiv p/\rho$ which contrarily to section 3, is not 
assumed here to be constant. Note that the metric is well defined for 
$\kappa^2_{(5)} b_0\rho < 1$.

To specify completely our solution, we need to determine the functions 
$a_0(t)$ and $\rho(t)$ (note that $n_0(t)$ is a totally arbitrary function 
whose choice represents simply a choice of time coordinate). This can 
be done by substituting  the metric ansatz 
(\ref{exactsolution}-\ref{lambdarho}) in the Einstein equations.  
The component $\tilde G_{00}=\kappa^2_{(5)}\tilde T_{00}$ taken on 
the brane, then yields  the equation
\beq
{{\dot a_0}^2\over a_0^2}=\frac{\kappa^4_{(5)}}{36}\rho^2. \label{H2}
\eeq
This equation, in addition to the energy conservation equation on the 
brane, 
\beq
\dot\rho+3{\dot a_0\over a_0}(1+w)\rho=0, \label{conserv}
\eeq
completely determines, once an equation of state has been specified, the 
unknown functions $a_0(t)$ and $\rho(t)$.

In the particular case of a constant $w$, 
the energy density conservation on the brane 
implies $\rho\propto a_0^{-3(1+w)}$ and (\ref{H2}) can then be integrated 
explicitly so that 
\beq
a_0(t)\propto t^q,\quad \kappa^2_{(5)} \rho={6q\over t}, 
\quad q={1\over 3(1+w)} \qquad (w\neq -1), 
\eeq
or for  a cosmological constant on the brane,
\beq
a_0(t)\propto e^{Ht}, \quad H={\kappa^2_{(5)}\over 6} \rho 
\qquad (w= -1).
\eeq
We have kept only the expanding solutions of (\ref{H2}).

 In the general case, 
it remains to   check that the metric so obtained 
(\ref{exactsolution}-\ref{lambdarho}) satisfies  
 {\it all}  Einstein equations  {\it everywhere}. This can indeed 
be done explicitly  using  (\ref{H2}) and (\ref{conserv}).

An interesting consequence of our solution is the behaviour of the second
brane. As in the previous section, the matter content of the second brane 
is totally determined by the ``topological constraints'' due to the compactness
of the additional dimension. Since our solution is a particular case of the 
general form (\ref{aglobal}), the same relations (\ref{rhoconstr}) and
(\ref{pconstr}) apply, which implies 
\beq
\rho_*=-\rho\left(1-\frac{\kappa^2_{(5)}}{12} b_0\rho\right)^{-1},
\label{rhorhostar}
\eeq
and 
\beq
w_*+{2\over 3}=\left(w+{2\over 3}\right){1-{\kappa_{(5)}^2\over 12}b_0\rho 
\over 1+{\kappa_{(5)}^2\over 4}b_0\rho  (w+2/3)}, \label{wstar}
\eeq
 where $w_*\equiv p_*/\rho_*$. 
In the general case, $w$ and  $w_*$ will be time-dependent. Even if 
$w$ is chosen to be constant, $w_*$ will be time-dependent, except in two 
particular cases:                
  $w=w_*=-1$ corresponding to  a cosmological constant on both branes
(although of opposite signs), and  $w=w_*=-2/3$.
Note that the role of the two branes is perfectly symmetric, as it is clear
from (\ref{rhoconstr}) and (\ref{pconstr}). It is not manifest on the two 
equations above, (\ref{rhorhostar}) and (\ref{wstar}), because we have 
privileged the position $y=0$ in the form 
(\ref{exactsolution}-\ref{linearmetric}) given to the metric. 

It can be checked explicitly that the energy conservation on the second 
brane, {\it i.e.} 
\beq
\dot \rho_*+3{\dot a_{1/2}\over a_{1/2}}(1+w_*)\rho_*=0,
\eeq
is automatically satisfied upon using  the first brane energy conservation 
 (\ref{conserv})  and the two  relations (\ref{rhorhostar}) 
and (\ref{wstar}). 

The present class of solutions could  be considered as an 
illustration of a holographic principle, by which
all the properties of the bulk (here the metric) can be determined
solely by the specification of the fields on the boundary (here the
metric and the matter in our brane-universe). 

Finally,
it is interesting to note that this new class of exact solutions includes as 
a particular case the solution given in \cite{kl98}. The latter solution 
corresponds to matter in the brane with the equation of state $p=-\rho$
and  represents  an 
extension of an earlier  domain wall inflationary solution 
(in a four-dimensional
spacetime) \cite{domainwall}.

\subsection{Kaluza-Klein approach}
 The geometrical picture of  section 4 showed us that the 
$y$-dependence of the five-dimensional metric, embodied in the extrinsic 
curvature tensor, is unavoidable because of the very presence of our 
brane-universe. The purpose of this section is to show how 
this picture 
can be reconciled with the seemingly contradictory Kaluza-Klein approach, 
for which, at an energy scale lower than the inverse radius of the fifth 
dimension, the so-called massive Kaluza-Klein modes (or in a more geometrical 
language, the $y$-dependence of the fields) can be ignored. 

The traditional Kaluza-Klein approach, whose purpose is to give a 
four-dimensional interpretation of the five-dimensional world, is supposed 
to work when the  energy scale  of the system, defined more specifically
as the typical inverse length scale or time scale of variation of the 
fields, is much smaller than the inverse of the radius 
of the fifth dimension, defined by 
\beq
R_{(5)}(\tau) = \int_{-1/2}^{1/2} b(\tau,y) dy.
\eeq
Since we are interested by cosmology, only time variations of the fields 
will be considered from a four-dimensional point of view. And the 
corresponding (inverse)
time scale is given by the Hubble parameter 
$H(\tau,y)=\dot{a}/ (an)$, 
so that the limit of validity of a Kaluza-Klein reduction is 
\beq
H R_{(5)}\ll 1. \label{HR5}
\eeq
For energy scales of the order or above $R_{(5)}^{-1}$, the fifth dimension is 
expected to ``open up".

The condition $HR_{(5)} \ll 1$ can also be reexpressed in terms of
 $\rho$ using (\ref{Fried}), in which $H$ is typically of the order 
of $\kappa^2_{(5)}\rho$.  The low energy limit should thus be valid when 
$\kappa^2_{(5)}\rho \ll R^{-1}_{(5)}$.\footnote{Note that this is also  the condition found
in \cite{low99} when one requires  that
the $|y|$-dependent part of the metric (given by $[a]'_0a_0 |y|$ to
the first order) should be small with respect to $a_0^2$, using equation
(\ref{aarho}).}

 In the traditional  way of thinking, it is expected that, 
 when $HR_{(5)} \rightarrow 0$, a given solution of
the equations of motion will converge towards its average value 
(or "zero-mode") over the compact space,  the average value for any 
quantity $Q$ being defined by
\begin{equation}
\langle Q \rangle= \int_{-\frac{1}{2}}^{\frac{1}{2}} dy  \, Q(y).
\end{equation}
This would mean in particular for the metric  that 
 $ g_{\mu\nu}(y,\tau) \rightarrow \langle g_{\mu\nu} \rangle(\tau)$ for
any $y$ and 
$\tau$, and especially for $y=0$ (on the brane) 
\footnote{From a purely mathematical point of view, one could imagine a
situation where the convergence would not be true at any point of spacetime, 
but  {\it almost
everywhere}; so that the dynamics of the
zero modes could be different from the one of the {\it induced metric}. 
This will not be the case for the solutions explicitly studied in this work.}.
 If this limit is true, then it is equivalent to obtain the
cosmological evolution from the equations of motion verified by the zero
modes or from the equations of motion satisfied by the metric of the brane.
In this spirit, it would seem natural to average the Einstein equations
over the fifth dimension.

The main difficulty, however, in this procedure is that (when  one neglects 
 bulk matter) the ``topological constraints'' mentioned previously will impose
the presence of a second brane, so that when one takes the average of 
Einstein's equations, for example the (0,0) component, one will obtain 
on the right hand side not only a term proportional to $\rho$ (which,
 were it be alone, would give back the standard Friedmann equations) but also 
a second term proportional to $\rho_*$. Moreover, as it can be seen 
explicitly with our exact solutions,  $\rho$ and $\rho_*$ ``conspire'' so that 
their sum will be proportional to $\rho^2$.

After the results of sections 3 and 4, this conclusion should not be 
surprising.  Indeed, 
  the non-standard features of the cosmological  
equations were shown, through the Gauss-Codacci formalism, to come from 
a simple geometrical interpretation, {\it i.e.} the embedding of our brane-universe
in the five-dimensional spacetime.  
This geometrical effect was obtained by a  purely local analysis
so that one sees no reason why it would cease to be valid when 
we "compactify" the theory, {\em i.e.} when we assume (\ref{HR5}).

Let us now try to be more specific and to exhibit, in the simple case
$b_0 = b_{1/2}$ (and thus, from (\ref{aprimconstr}), $\jl a' \jr_0  = 
- \jl a' \jr_{1/2}$), a decomposition of
the fields into zero modes and massive modes as in the Kaluza-Klein 
approach. We take a standard  decomposition for the function $b$ 
in the form 
\beq
b=B_{0}+\sum_{k \neq 0} B_{k}e^{2 i \pi k y}, \label{bKK}
\eeq
where $B_0\equiv \langle b\rangle$ is the zero mode. The analogous 
decomposition in the case of $a$ and $n$ is more subtle because 
they must {\it necessarily} have a dependence 
on $y$ due to the presence of the brane. 
This means that a metric solution of the five-dimensional Einstein 
equations, because of the presence  of a $y$-dependent source term, 
will be necessarily $y$-dependent in order to keep 
the compactification  procedure consistent (see e.g. \cite{dnpw84}).
To make this property manifest,
we choose to adopt a distinction between what we 
call ``branic'' modes and the classical Kaluza-Klein modes, and write
\begin{eqnarray}
n&=&N_{0}
+\sum_{k \neq 0}\left(-{\aleph_n\over 4\pi^2 k^2}(1-(-1)^k) 
+ N_{k}\right) e^{2 i \pi k y}, \\
a&=&A_{0}+\sum_{k \neq 0} \left(-{\aleph_a\over 4\pi^2 k^2}(1-(-1)^k)
 + A_{k}\right) e^{2 i \pi k y}. \label{aKK}
\end{eqnarray}
In the expressions above, the terms proportional to $\aleph$ 
correspond to the {\it branic modes}. 
According to (\ref{asec}), $\aleph_a=[a']$ and 
$\aleph_n=[n']$. The ``branic'' terms in the decomposition come from the 
Fourier transform of $|y|$. 
The ``branic" modes  are different from the classical 
Kaluza-Klein massive modes in the sense that their effective mass is 
independent 
of $k$.
The expansions above imply that $\langle n\rangle=N_0$
and $\langle a\rangle=A_0$.

It is now instructive to compare the zero-mode metric and the ``true'' 
metric on the brane in the case of our explicit solutions of the  previous 
subsection. The average over the fifth dimension of $a$ and $n$ yields 
respectively 
\beq
A_0=a_0\left(1-\frac{\kappa^2_{(5)}}{24} b_0\rho\right), \qquad
N_0=n_0\left(1+\frac{\kappa^2_{(5)}}{24}(3w+2) b_0\rho\right).
\eeq
The first remark is that the zero-modes of the metric indeed converge 
towards the ``true'' metric on the brane (which is the meaningful one 
as far as observations are concerned) in the limit 
$\kappa^2_{(5)}\rho \ll R^{-1}_{(5)}$ (here $R_{(5)}=b_0$).
The same will be true for the time variation of the metric. 
The ``Kaluza-Klein'' Hubble parameter is indeed related to the ``true'' Hubble 
parameter by the expression
\beq
{\dot A_0\over N_0 A_0}={\dot a_0\over n_0 a_0}
\left(1-\frac{\kappa^2_{(5)}}{24} b_0\rho\right)^{-1}.
\eeq

Similar conclusions can be obtained from a variation principle for the 
case of a cosmological constant (for which the action is well defined). 
Starting from the action
\begin {equation}
S= \int d^4xdy \sqrt{-\tilde{g}} \left( -\frac{1}{2\kappa^2_{(5)}} \tilde{R}+
  \frac{\delta(y)}{\sqrt{\tilde{g}_{55}}} \rho +
  \frac{\delta(y-1/2)}{\sqrt{\tilde{g}_{55}}} \rho_* \right),
\eeq
where $\rho$ and $\rho_*$ are constants, 
one can then obtain the equation of motion for the zero modes by a naive compactification, taking into
account the constraint (\ref{rhorhostar}). We get, 
after compactification
\begin {equation}
S= R_{(5)} \int d^4x \sqrt{-g} \left( -\frac{R}{2\kappa^2_{(5)}}  
- \frac{\kappa^2_{(5)}}{12}\rho^2
   \right).
\eeq
This gives  the following Friedmann-type  equations:
\beq
3\frac{\ddot{A_0}}{A_0} = \frac{\kappa_{(5)}^4}{12}\rho^2,\quad 
\frac{\ddot{A_0}}{A_0} + 2\frac{\dot{A_0}^2}{A_0^2}=
\frac{\kappa_{(5)}^4}{12}\rho^2,
\eeq
which show  that $A_0$ obeys the same equation (\ref{Fried}) obeyed
 by $a_0$ (with $\rho=-p$ and $\check{T}_{55}=0$). 

To summarize, the Kaluza-Klein approach remains valid in a ``branic'' 
context but one must be careful and take into account the ``topological 
constraints'' in the average procedure over the matter sources. One then 
recovers the non-standard equations obtained in the previous sections from 
a local point of view.

\section{Cosmological consequences}
The main purpose of this work was to show that, quite generically, 
the cosmology arising from the assumption that our universe is confined 
to a three-brane within a larger spacetime {\it does not lead to the 
Friedmann equations of standard cosmology}. The most striking features of the 
new equations are on one hand that $H^2$ is not proportional to $\rho$ as in 
standard cosmology but rather is related to an expression quadratic in $\rho$,
on the other hand that the effective four-dimensional Newton's constant
$G_{(4)}$
does not enter, in other words the equations do not depend on the 
radius of the fifth dimension but only on the five-dimensional Newton's
constant $G_{(5)}$, {\em i.e.} the  Planck mass scale of the fundamental 
theory.

One of the important questions is whether such a cosmology is viable 
when confronted to observations. Let us discuss the consequences when we
assume that equation (\ref{Fried}) applies
to our present universe.  
It is convenient to rewrite the  energy density 
$\rho$ as a fraction of the critical density $\rho_c$ whose
usual definition is
\beq
\rho_c={3 H^2\over 8\pi G_{(4)}},
\eeq
such that 
\beq
\rho= \Omega \rho_c.
\eeq
One can now rewrite 
(\ref{Fried}) in the form 
\beq
\dot H+2H^2=-{1\over 4}\lambda^2H^2 \Omega^2\left(1+3w\right)
-{1\over 3} \kappa_{(5)}^2 \check{T}_5^5 , \label{today}
\eeq
where $\lambda$ is a dimensionless parameter defined by
\beq
\lambda={m_p^2 H \over M_{(5)}^3}, 
\eeq
 having introduced the  reduced four-dimensional Planck mass $m_p=\kappa_{(4)}^{-1}$, 
 which can be given in terms of the fundamental scale as
 \begin{equation}
 m_p^2=\left(8\pi G_{(4)}\right)^{-1}=M_{(5)}^3 R_{(5)}.
 \end{equation} 
Thus, $\lambda$ can also be interpreted as the ratio of the fifth dimension 
radius over the  Hubble radius. 
The left hand side of the above equation (\ref{today}) 
can be reexpressed today 
in terms of the so-called acceleration parameter (or rather deceleration 
parameter) defined by 
\beq
q=-\left({\ddot a a\over \dot a^2}\right)_{now},
\eeq 
so that 
\beq (\dot H+2H^2)_{now}=(1-q)H_{now}^2.
\eeq
Cosmological observations constrain  $|q|$ to be of the order $1$.
By comparing the two sides of equations (\ref{today})
($\Omega_{now}$ being of the order $1$),  one thus arrives 
to the conclusion that the present matter content of our universe could
influence the dynamics of its geometry if $\lambda$ was at least of order $1$, 
which would signify that the fifth dimension radius is of the order of the 
present Hubble radius. Such a possibility, however, must 
 be rejected because  we could then 
observe directly five-dimensional gravity, which is 
not the case. As we discuss below, a possible way out is to assume that
the bulk term dominates in equation 
(\ref{Fried})  and is driving the dynamics of our universe.

However, another possibility is that equation (\ref{Fried}) does not
apply today. Indeed, an essential assumption of our model is strict
homogeneity and isotropy of our Universe which is only true today on very
large scales. Therefore our model probably breaks down when deviations
from homogeneity are important on scales of the order of the fifth
dimension radius.

One could then consider the possibility that this non-conventional 
cosmology existed only 
for the early universe. How would it affect nucleosynthesis ?
Although detailed constraints would make necessary the use of a numerical 
evolution of the chemical reactions, one can already get a rough constraint 
from the abundance of Helium by taking the approximation 
that all free neutrons just before nucleosynthesis will end up 
in Helium nuclei.
To see what this would entail in our unconventional scenario, let us 
recall that nucleosynthesis can be roughly divided into two steps: 
the first step consists in the freezing-out of the proton to neutron ratio, 
which  takes place at a temperature $T\simeq T_D$, 
the second step is the production of light nuclei, which occurs at 
a temperature $T\simeq T_N \sim 0.1 \mbox{MeV}$ (we neglect here the finite time span for 
this step).  

In the standard scenario (see, e.g., \cite{weinberg}), 
above the temperature $T_D\simeq
(0.7-0.8)$ MeV, the neutron to proton ratio is fixed by the thermodynamic
equilibrium relation 
\beq
\left({n\over p}\right)_{eq}(T)=\exp\left(-Q/T\right),
\eeq
where $Q\equiv m_n-m_p \simeq 1.3 \mbox{MeV}$. 
Below $T_D$, as long as the production of light nuclei has not begun, 
the neutron to proton ratio will be given by  $(n/p)_{eq}(T_D)$ corrected 
by the decay of neutrons into protons (occuring on the time scale 
$\tau_n\sim 10^3$ s), so that 
\beq
\left({n\over p}\right)(T)\simeq 
\left({n\over p}\right)_{eq}(T_D)\exp(-t/\tau_n),
\eeq
where $t$ is the age of the universe at the temperature $T$ 
(assuming $t \gg t_D$).
Assuming all neutrons existing at $T_N^{}$, just before 
production of light nuclei,  will be essentially transformed into 
Helium nuclei, one gets $Y=2x/(1+x)$ where $Y$ is the mass fraction 
of Helium 4 and $x$ is the initial 
neutron to proton ratio. Observations give us $Y\simeq 0.25$, which implies
$x=Y/(2-Y)\simeq 1/7$.

In the unconventional scenario (assuming the bulk 
contribution is negligible at the time of nucleosynthesis), 
$a(t)\propto t^{1/4}$ so that 
the age of the universe is $t\simeq H^{-1}/4$. Moreover $H\propto T^4$
(instead of $H\propto T^2$ in the standard model), which 
implies that  the cooling in the universe is much slower than in the standard 
case. An immediate consequence is that $T_D$ cannot be the same as in the
standard scenario, because it would lead to a much smaller neutron to proton 
ratio just before nucleosynthesis due to a more effective decay of neutrons 
into protons. So let us now evaluate the required new value for $T_D$.
For $T\gg Q$, the rate $\lambda_{np}$ of the reaction 
$n\leftrightarrow p$ is given by 
\beq
\lambda_{np}=\alpha T^5 ,
\eeq
with $\alpha\simeq 0.76 s^{-1}{\rm MeV}^{-5}$. 
Let us also write the Hubble parameter in the form
\beq
H=\beta T^4 . \label{hbeta}
\eeq
The freezing-out temperature $T_D$ is reached when the `chemical' reactions 
slow down to the cosmological 
expansion rate, {\it i.e.} when $H\simeq \lambda_{np}$, 
and is therefore given by $T_D\simeq \beta/\alpha$. It thus follows
 that $\beta$ can be expressed  in the form 
\beq
\beta \simeq -\left( \alpha  Q +{1\over 4} T_N^{-4}
\tau_n^{-1}\right)/ \ln x.
\eeq
Inserting numerical values, one finds $T_D\simeq 2-3$ MeV. More importantly,
 the scale of $M_{(5)}$ is also fixed by this constraint, substituting 
the value of $\beta$ in (\ref{hbeta}) and then $H$ in (\ref{Fried}) 
with $\rho=3p=(\pi^2/30) g_*T^4$ (where $g_*$ is the effective 
number of relativistic degrees of freedom). One then finds
\beq
 g_*^{-\frac{1}{3}}M_{(5)} \simeq \left( \frac{\pi^2}{180 \beta} 
\right)^{\frac{1}{3}} \sim 3 \, \mbox{TeV}.
\eeq
For five dimensions this fundamental scale is too low to be in agreement with 
gravity such as it appears to us on large scales, but it would be worthwhile 
to study the cases with more dimensions.

Going (much) further into the past, one can examine  the issue of
inflation in brane cosmology, a question which has attracted recently 
much attention \cite{l99}-\cite{ko99}. 

A first question is whether inflation is triggered by the energy density
in the bulk or in the brane. It is more in the spirit of this paper to
discuss brane inflation and we will restrict our attention to this
case. The next issue deals with the initial conditions. 
It has been stressed that brane inflation might require an almost empty
bulk, following the condition $\rho_{_{\rm bulk}} R_{(5)} \ll 
\rho_{_{\rm brane}}$. The study above clearly shows that the domination
of brane over bulk energy density need only be imposed at a local level,
{\em i.e.} on a typical distance scale of order $M_{(5)}^{-1}$. In other
words, we need only to impose the weaker constraint 
\beq
\rho_{_{\rm bulk}}
M_{(5)}^{-1} <  \rho_{_{\rm brane}}. \label{dombraneapp}
\eeq
This in fact ensures that the
bulk energy in the neighbourhood of the brane does not dominate over the
confined energy. It is also consistent with our general constraint 
(\ref{dombrane}), under the condition $ \rho_{_{\rm brane}} <
M_{(5)}^4$, which represents a natural bound for the energy density
localized in the brane. 

Similarly, recent discussions \cite{bd98,l99,kl98} 
on the necessity of having an untolerably
light inflaton field do not apply in the context of our non-conventional
cosmological scenario. Indeed, using our relation (\ref{Fried}) and 
assuming the natural 
bound $\rho\le M^4_{(5)}$, we obtain
\beq
H\sim M^{-3}_{(5)}\rho \leq M_{(5)}.
\eeq
This only constraints the inflaton mass $m$, which must be less than 
$H$ during inflation in order to fulfil slow-roll conditions, to be 
less than the fundamental scale $M_{(5)}$.

Throughout this work, we have focused our attention to the cases where 
the matter content of the brane dominates that of the bulk. Let us 
however point out that, if the bulk was dominating, then
one would be back in the traditional Kaluza-Klein approach, with the
difference that invisible five-dimensional matter would control the
geometry of the Universe whereas ordinary matter would play only a
passive r\^ole. However, an important constraint on such scenarios is that they
must recover the remarkable agreement, in standard cosmology,
 of the present values of the
Hubble parameter and of the radiation temperature with their respective
values at the time of nucleosynthesis, {\em i.e.} on a span of almost
ten orders of magnitude.

Finally, a word should be said about the difficult problem of the
stabilisation of the compact dimension \cite{add98,adm98b,ddgr99}. Our previous
considerations rest in a crucial way on the assumption of invariance
under diffeomorphisms of the five-dimensional spacetime. The
stabilisation of the radius $R_{(5)}$ is usually obtained 
 with the introduction
of a potential for the corresponding $b$ scalar field, which obviously
breaks this reparametrisation invariance. Our formalism would thus not be
applicable directly once the radius has been stabilized. On the other
hand, this potential usually arises through quantum or thermal 
fluctuations of the five-dimensional theory. In this case, the
five-dimensional framework that we use should be applicable, although our
considerations have clearly been limited to a classical point of view.

It would be interesting to develop the present work in several
directions.
First, as we mentioned earlier, our model does not necessarily apply to
recent cosmology because of deviations from homogeneity. An important
extension would thus be to allow for perturbations in the
brane-universe. Another direction is to allow for more than one extra 
dimension. Finally, in the case of string models, the presence of scalar
fields such as the dilaton or Ramond-Ramond scalars may play a r\^ole
in the early Universe dynamics.

\vskip 1cm
{\bf Acknowledgments}
\vskip .3cm
We wish thank Brandon Carter, Emilian Dudas and Jihad Mourad for 
fruitful discussions.


\end{document}